\documentclass[useAMS,usenatbib]{mn2e}
\usepackage{amsmath}
\usepackage{amssymb}
\usepackage{graphicx}
\usepackage{subfig}
\usepackage{epsfig}
\usepackage{caption}
\newcommand*{\bigchi}{\mbox{\Large$\chi$}}

\title[The Ratio of CO to Total Gas Mass in High Redshift Galaxies]{The Ratio of CO to Total Gas Mass in High Redshift Galaxies}
\author[N. Mashian et al.]{Natalie Mashian$^{1,2}$\thanks{nmashian@physics.harvard.edu}, Amiel Sternberg$^{2}$\thanks{amiel@wise.tau.ac.il}, Abraham Loeb$^{1,2}$\thanks{aloeb@cfa.harvard.edu} \\
$^{1}$Harvard-Smithsonian Center for Astrophysics, 60 Garden Street, Cambridge, MA 02138, USA\\
$^{2}$The Raymond and Beverly Sackler School of Physics and Astronomy, Tel Aviv University, Tel Aviv 69978, Israel}

\begin{document}

\pagerange{\pageref{firstpage}--\pageref{lastpage}} \pubyear{2013}

\maketitle

\label{firstpage}


\begin{abstract}
\citet{walt} have recently identified the J=6-5, 5-4, and 2-1 CO rotational emission lines, and [C{\scriptsize II}] fine-structure emission line from the star-forming interstellar medium in the high-redshift submillimeter source HDF 850.1, at z = 5.183. We employ large velocity gradient (LVG) modeling to analyze the spectra of this source assuming the [C{\scriptsize II}] and CO emissions originate from \emph{(i)} separate virialized regions, \emph{(ii)} separate unvirialized regions, \emph{(iii)} uniformly mixed virialized regions, and \emph{(iv)} uniformly mixed unvirialized regions. We present the best fit set of parameters, including for each case the ratio $\alpha$ between the total hydrogen/helium gas mass and the CO(1-0) line luminosity. We also present computations of the ratio of H$_2$ mass to [C{\scriptsize II}] line-luminosity for optically thin conditions, for a range of gas temperatures and densities, for direct conversion of  [C{\scriptsize II}]  line-luminosities to `` CO-dark" H$_2$ masses. For HDF 850.1 we find that a model in which the CO and C$^+$ are uniformly mixed in gas that is shielded from UV radiation, requires a cosmic-ray or X-ray ionization rate of $\zeta \approx$ 3$\times$ 10$^{-14}$ s$^{-1}$, plausibly consistent with the large star-formation rate ($\sim$ 10$^3$ M$_{\odot}$ yr$^{-1}$) observed in this source. Enforcing the cosmological constraint posed by the abundance of dark matter halos in the standard $\Lambda$CDM cosmology and taking into account other possible contributions to the total gas mass, we find that the two models in which the virialization condition is enforced can be ruled out at the $\gtrsim$ 2$\sigma$ level while the model assuming mixed unvirialized regions is less likely. We conclude that modeling HDF 850.1's ISM as a collection of unvirialized molecular clouds with distinct CO and C$^+$  layers, for which $\alpha$ = 1.2 M$_{\odot}$ (K km s$^{-1}$ pc$^2$)$^{-1}$ for the CO to H$_2$ mass-to-luminosity ratio, (similar to the standard ULIRG value), is most consistent with the $\Lambda$CDM cosmology.
\end{abstract}


\begin{keywords}
cosmology: theory -- galaxies: high-redshift -- galaxies: ISM -- galaxies: mass function.
\end{keywords}

\section{Introduction}

Observations of high-redshift CO spectral line emissions have greatly increased our knowledge of galaxy assembly in the early Universe. At redshifts z$\sim$2, this includes the discovery of turbulent star-forming disks with cold-gas mass fractions and star-formation rates significantly larger than in present day galaxies \citep{daddi, genzel, mag, tacc, taccnopub}. The high star-formation rates are correlated with large gas masses and luminous CO emission lines \citep{ken} observable to very high redshifts. A prominent example is the luminous submillimeter and Hubble-Deep-Field source HDF 850.1, which is at a redshift of z = 5.183 as determined by recent detections of CO(6-5), CO(5-4), and CO(2-1) rotational line emissions, and also [C{\scriptsize II}] fine-structure emission in this source \citep{walt}. The high redshift of HDF 850.1 offers the opportunity of setting cosmological constraints on the conversion factor from CO line luminosities to gas masses, via the implied dark matter masses and the expected cosmic volume density of halos of a given mass. Such analysis is the subject of our paper.
\\ \indent  CO emitting molecular clouds, which provide the raw material for star formation, are usually assumed to have undergone complete conversion from atomic to molecular hydrogen. However, since H$_2$ has strongly forbidden rotational transitions and requires high temperatures ($\sim$500 K) to excite its rotational lines, it is a poor tracer of cold ($\lesssim$ 100 K) molecular gas. Determining H$_2$ gas masses in the interstellar medium (ISM) of galaxies has therefore relied on tracer molecules. In particular, $^{12}$CO is the most commonly employed tracer of ISM clouds; aside from being the most abundant molecule after H$_2$, CO has a weak dipole moment ($\mu_{e}$ = 0.11 Debye) and its rotational levels are thus excited and thermalized by collisions with H$_2$ at relatively low molecular hydrogen densities \citep{sovan}.
\\ \indent The molecular hydrogen gas mass is often obtained from the CO luminosity by adopting a mass-to-luminosity conversion factor $\alpha = M_{H_2}/L'_{CO(1-0)}$ between the H$_2$ mass and the J =1-0 115 GHz CO rotational transition \citep{bol}. The value for $\alpha$ has been empirically calibrated for the Milky Way Galaxy by three independent techniques: \emph{(i)} correlation of optical extinction with CO column densities in interstellar dark clouds \citep{dick}; \emph{(ii)} correlation of gamma-ray flux with the CO line flux in the Galactic molecular ring \citep{blo, str}; and \emph{(iii)} observed relations between the virial mass and CO line luminosity for Galactic GMCs \citep{solo}. These methods have all arrived at the conclusion that the conversion factor in our Galaxy is fairly constant. The standard Galactic value is $\alpha$ = 4.6 M$_{\odot}$/(K km$^{-1}$ pc$^{2}$). Subsequent studies of CO emission from unvirialized regions in Ultra-Luminous Infrared Galaxies (ULIRGs) found a significantly smaller ratio. For such systems, $\alpha$ = 0.8 M$_{\odot}$/(K km$^{-1}$ pc$^{2}$) \citep{doso}. These values have been adopted by many \citep{san, tin, wang, walt} to convert CO J=1-0 line observations to total molecular gas masses, but without consideration of the dependence of $\alpha$ on the average molecular gas conditions found in the sources being considered. Since all current observational studies of $\alpha$ leave its range and dependence on the average density, temperature, and kinetic state of the molecular gas  still largely unexplored, its applicability to other systems in the local or distant Universe is less certain \citep{pap}.
\\ \indent In this paper, we estimate the CO emitting gas masses in HDF 850.1 using the large-velocity-gradient (LVG) formalism to fit the observed emission line spectral energy distribution (SED) for a variety of model configurations, and for a comparison to the Galactic and ULIRG conversions. In \S2 we outline the details of the LVG approach, including an overview of the escape probability method and a derivation of the gas mass from the line intensity of the modeled source. In \S3, we present the best fit set of parameters that reproduce HDF 850.1's detected lines and calculate the corresponding molecular gas mass, assuming the CO and [C{\scriptsize II}] emission lines originate from \emph{(i)} separate virialized regions, \emph{(ii)} separate unvirialized regions, \emph{(iii)}uniformly mixed virialized regions, and \emph{(iv)} uniformly mixed unvirialized regions. The inferred gas masses enable us to set lower-limits on the dark-matter halo mass for HDF 850.1. In \S4 we compare the estimated halo-masses to the number of such objects expected at high redshift in $\Lambda$CDM cosmology, and show that models with lower values of $\alpha$ are favored. We conclude with a discussion of our findings and their implications in \S5.

\section{Large Velocity Gradient Model}
We start by describing our procedure for quantitatively analyzing the [C{\scriptsize II}] and CO emission lines detected at the position of HDF 850.1 using the large velocity gradient (LVG) approximation. We consider a multi-level system with population densities of the \emph{i}th level given by \emph{n$_{i}$}. The equations of statistical equilibrium can then be written as :
\begin{equation}
n_{i}\sum_{j\neq i}^{l} R_{ij} = \sum_{j\neq i}^{l} n_{j}R_{ji}
\end{equation}
where \emph{l} is the total number of levels included; since the set of \emph{l} statistical equations is not independent, one equation may be replaced by the conservation equation
\begin{equation}
n_{tot}=\sum_{j=0}^{l}n_{j}
\end{equation}
where \emph{n$_{tot}$} is the number density of the given species in all levels. In our application, $n_{tot}$ = $n_{CO}$.
Following the notation of \citet{poel}, \emph{R$_{ij}$} is given in terms of the Einstein coefficients, \emph{A$_{ij}$} and \emph{B$_{ij}$}, and the collisional excitation (\emph{i $<$ j})and de-excitation rates (\emph{i $>$ j})  \emph{C$_{ij}$}: 
\begin{equation}
R_{ij}=
\begin{cases}
	A_{ij}+B_{ij}\langle J_{ij}\rangle+C_{ij},& (i > j) \\
	B_{ij}\langle J_{ij}\rangle+C_{ij}, 	       & (i < j)
\end{cases}
\end{equation}
where $\langle J_{ij}\rangle$ is the mean radiation intensity corresponding to the transition from level \emph{i} to \emph{j} averaged over the local line profile function $\phi(\nu_{ij})$. The total collisional rates $C_{ij}$ depend on the individual temperature-dependent rate coefficients and collision partners, usually H$_2$ and H for CO rotational excitation.  
\\ \indent The difficulty in solving this problem is that the mean intensity at any location in the source is a function of the emission and varying excitation state of the gas all over the rest of the source, and is thus a nonlocal quantity. To obtain a general solution of the coupled sets of equations describing radiative transfer and statistical equilibrium, we adopt the approach developed by \citet{sobo} and extended by \citet{cast} and \citet{lucy} and assume the existence of a large velocity gradient in dense clouds. This assumption is justified given the interstellar molecular line widths which range from a few up to a few tens of kilometers per second, far in excess of plausible thermal velocities in the clouds \citep{gold}. They suggest that these observed velocity differences arise from large-scale, systematic, velocity gradients across the cloud, a hypothesis that lies in accord with the constraints provided by observation and theory. 

\begin{table*}
 \begin{minipage}{140mm}
  \caption{CO and [C{\scriptsize II}] Line Observations of HDF 850.1}
  \begin{tabular}{@{}llrrrrlrlr@{}}
  \hline
   Line & $\nu_{obs}$
	& Integrated Flux Density
	& Luminosity & Intensity  \\
&  $\rm{[GHz]}$ & $\rm{[Jy\: km\: s^{-1}]}$ & $[10^{10}\: \rm{K\: km\: s^{-1}\: pc^2]}$
	& $\rm{[erg(s\: cm^2\: str)^{-1}]}$ \\
 \hline
CO(2-1) & 37.286 & 0.17$\pm$0.04 & 4.1$\pm$0.9 & (2.16$\pm$0.51)$\times$10$^{-9}$  \\
CO(5-4) & 93.202 & 0.5$\pm$0.1 & 1.9$\pm$0.4 & (1.59$\pm$0.32)$\times$10$^{-8}$\\
CO(6-5) & 111.835 & 0.39$\pm$0.1 & 1.0$\pm$0.3 & (1.49$\pm$0.82)$\times$10$^{-8}$\\
{[}C{\scriptsize II}{]}  & 307.383 & 14.6$\pm$0.3 & 5.0$\pm$0.1 & (1.5$\pm$0.03)$\times$10$^{-6}$\\
\hline
\end{tabular}
\\
Detection of four lines tracing the star-forming interstellar medium in HDF 850.1 \citep{walt}	
\end{minipage}
\end{table*}

In the limit that the thermal velocity in the cloud is much smaller than the velocity gradient across the radius of the cloud, the value of $\langle J_{ij}\rangle$ at any point in the cloud, when integrated over the line profile, depends only upon the local value of the source function and upon the probability that a photon emitted at that point will escape from the cloud without further interaction. Thus  $\langle J_{ij}\rangle$ becomes a purely local quantity, given by:
\begin{equation}
\langle J_{ij}\rangle=(1-\beta_{ij})S_{ij}+\beta_{ij}B(\nu_{ij},T_{B})
\end{equation}
where $S_{ij}$ is the line source function,
\begin{equation}
S_{ij}=\frac{2h\nu_{ij}^3}{c^2}(\frac{g_in_j}{g_jn_i}-1)^{-1}
\end{equation}
assumed constant through the medium. In this expression $g_i$ and $g_j$ are the statistical weights of levels \emph{i} and \emph{j} respectively, $\beta_{ij}$ is the ``photon escape probability", and \emph{B$_{ij}(\nu_{ij}, T_{B}$)} is the background radiation with temperature $T_B$. In our models we set $T_B$ to the CMB temperature of 16.9 K at z = 5.183. We ignore contributions from warm dust \citep{cunha}.

 For a spherical homogenous collapsing cloud, the probability that a photon emitted in the transition from level \emph{i} to level \emph{j} escapes the cloud is given by
\begin{equation}
\beta_{ij}=\frac{1-e^{-\tau_{ij}}}{\tau_{ij}}
\end{equation}
where $\tau_{ij}$ is the optical depth in the line,
\begin{equation}
\tau_{ij}=\frac{A_{ij}}{8\pi}\frac{c^3}{\nu_{ij}^3}\frac{n_{tot}}{dv/dr}\frac{n_j g_{i}}{n_{tot}g_j}(1-\frac{g_j n_i}{g_i n_j}) .
\end{equation}
The equations of statistical equilibrium are therefore reduced to the simplified form
\begin{equation}
\sum_{j\neq i}^{l}(n_i C_{ij}-n_j C_{ji})+ n_i\sum_{j\neq i\atop j<i}^{l}A_{ij}\beta_{ij}-\sum_{j\neq i\atop j>i}^{l}n_jA_{ji}\beta_{ji}=0
\end{equation}
and can be solved through an iterative process to give the fractional level populations \emph{n$_i$/n$_{tot}$} (for a given choice of densities for the collision partners, usually H$_2$ but also H, and kinetic temperature \emph{T$_{kin}$}). Assuming the telescope beam contains a large number of these identical homogeneous collapsing clouds (e.g., \citealt{stev}), the corresponding emergent intensity of an emission line integrated along a line of sight is then simply
\begin{equation}
I_{ij}=\frac{h\nu_{ij}}{4\pi}\frac{n_i}{n_{tot}}A_{ij}\beta_{ij}N_{tot}=\frac{h\nu_{ij}}{4\pi}\frac{n_i}{n}A_{ij}\beta_{ij}\chi N
\end{equation}
where $\chi$ is the abundance ratio, $\chi \equiv n_{tot}/n$, where \emph{n} is the total hydrogen gas volume density (cm$^{-3}$) and \emph{N} is the hydrogen column density (cm$^{-2}$). 
\\ \indent Given a series of observed lines with frequency $\nu_{ij}$, one can identify a set of characterizing parameters that best reproduces the observed line ratios and intensity magnitudes; among these parameters are the cloud's kinetic temperature (\emph{T$_{kin}$}), velocity gradient (\emph{dv/dr}), gas density $n$ and collision partner (H and H$_2$) gas fractions, abundance ratio ($\chi$), and column density (\emph{N}). For a spherical geometry, this column density can then be further related to the molecular gas mass of the cloud in the following way
\begin{equation}
M_=\pi R^2 \mu m N' 
\end{equation}
where the factor $\mu=1.36$ takes into account the helium contribution to the molecular weight and $R=D_A\theta/2$ is the effective radius of the cloud, with $D_A$ being the angular diameter distance to the source and $\theta$ the beam size of the line observations. $N$ is defined in terms of the column density obtained from the LVG calculation, $N'=N(1+z)^4$ where the $(1+z)^4$ multiplicative factor reflects the decrease in surface brightness of a source at redshift \emph{z} in an expanding universe.
\\ \indent In the case where the emitting molecular clouds are gravitationally bound, applying the virial theorem to a homogenous spherical body yields the following constraint \citep{goldsmith},
\begin{equation}
\frac{dv}{dr}\approx \sqrt{\frac{G\pi \mu m n}{15}} \approx a\sqrt{\frac{n}{\rm{(cm^{-3})}}} \:\rm{km\: s^{-1}\: pc^{-1}}
\end{equation}
where \emph{a} = 7.77$\times$10$^{-3}$ if \emph{n} = \emph{n$_{H_2}$} and \emph{a} = 5.50$\times$10$^{-3}$ if \emph{n} = \emph{n$_{H}$} . The velocity gradient is inversely proportional to the dynamical time scale. In models where the clouds are assumed to be virialized, \emph{dv/dr} and \emph{n} are no longer independent input parameters of the model, but rather vary according to equation (11). 
\\ \indent For the optically thick $^{12}$CO J=1-0 transition  line ($\beta \approx 1/\tau$), carrying out the LVG calculations with this additional virialization condition leads to a simple relation between the gas mass and the  CO(1-0) line luminosity,
\begin{equation} 
\alpha_{CO}=\frac{M_{H_2}}{L'_{CO(1-0)}}=8.6\frac{\sqrt{n_{H_2}/(\rm{cm^{-3}})}}{T_{exc}/\rm{(K)}} \:\rm{{M_{\odot}}(K\: km\: s^{-1}\: pc^2)^{-1}}
\end{equation}
where the excitation temperature $T_{exc}\approx T_{kin}$ when the emission line is thermalized. (In this expression we assume complete conversion to H$_2$ so that the gas mass is the H$_2$ mass.) Empirically, the Galactic value of this mass-to-luminosity ratio for virialized objects bound by gravitational forces is $\alpha$ = M$_{H_2}$/L'$_{CO(1-0)}$ = 4.6 M$_{\odot}$ (K km s$^{-1}$ pc$^2$)$^{-1}$ \citep{solbar}, corresponding to $\sqrt{n_{H_2}}/T \sim$ 0.5 cm$^{-3/2}$K$^{-1}$.

\section{Analysis of HDF 850.1}

\begin{figure*}
\centerline{\includegraphics[width=400pt]{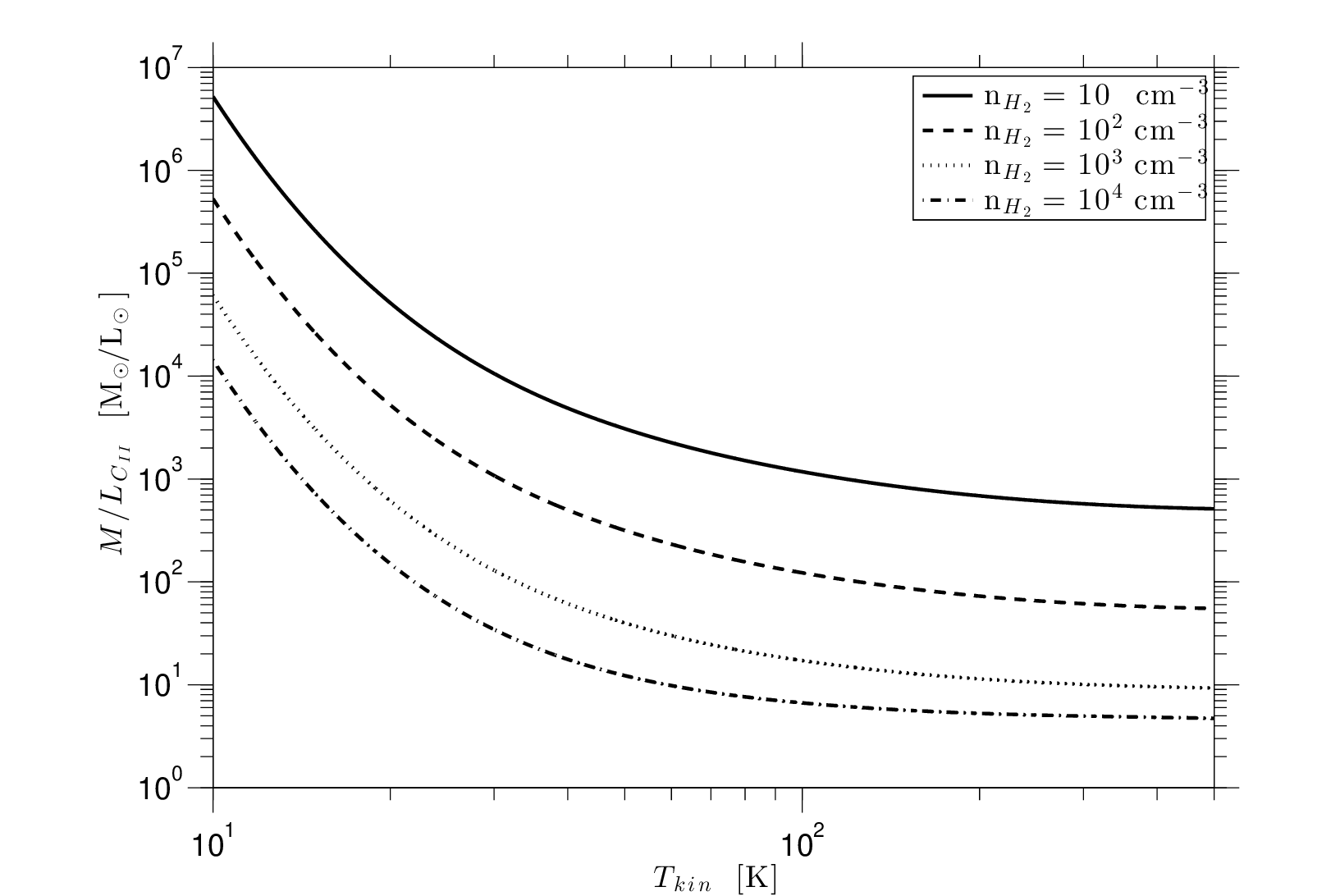}}
\caption{The gas mass to [C{\scriptsize II}] luminosity ratio for a C$^+$ to H$_2$ abundance ratio of $\bigchi_{C^+}$=10$^{-4}$ as a function of the kinetic temperature T$_{kin}$ at a molecular hydrogen number density of n$_{H_2}$ = 10, 10$^2$, 10$^3$, and 10$^4$ cm$^{-3}$. Calculations were made in the optically thin regime, assuming the fine-structure transition J=3/2$\rightarrow$1/2 is due solely to spontaneous emission processes and collisions with ortho- and para-H$_2$.}
\end{figure*}

\subsection{Properties of the Galaxy}
HDF 850.1, the brightest submillimetre source in the Hubble Deep Field at a wavelength of 850 micrometers, was discovered by \citealt{hughes}. A full-frequency scan of this source by \citet{walt} using the IRAM (Institut de Radioastronomie Millimetrique) Pleateau de Bure Interferometer and the National Radio Astronomy Observatory (NRAO) Jansky Very Large Array has detected three CO lines, identified as the CO(2-1) 230.5 GHz, CO(5-4) 576.4 GHz, and CO(6-5) 691.5 GHz rotational transitions. [C{\scriptsize II}] 1900.1 GHz (158 $\mu$m), one of the main cooling lines of the star-forming in stellar medium, has also been detected (Table 1). These lines have placed HDF 850.1 in a galaxy over density at z = 5.183, a redshift higher than those of most of the hundreds of submillimetre-bright galaxies identified thus far. 
\\ \indent \citet{walt} used an LVG model to characterize the CO spectral energy distribution of HDF 850.1 and found that the observed CO line intensities could be fit with a molecular hydrogen density of 10$^{3.2}$ cm$^{-3}$, a velocity gradient of 1.2 km s$^{-1}$pc$^{-1}$, and a kinetic temperature of 45 K. 
Then, assuming $\alpha$ = 0.8 M$_{\odot}$(K km s$^{-1}$pc$^{2}$)$^{-1}$ as for ULIRGs, they used the 1-0 line luminosity inferred from their LVG computation to infer that M$_{H_2}$ = 3.5$\times$10$^{10}$ M$_{\odot}$. However, as \citet{pap} argue, adopting a uniform value of $\alpha$ for ULIRGs neglect its dependence on the density, temperature, and kinematic state of the gas; this may limit the applicability of computed conversion factors to other systems in the local or distant Universe.
\\ \indent Here, we broaden the LVG analysis carried out in \citealt{walt} and use the LVG-modeled column density to estimate the total gas mass  of the source. We present several alternative models, each subject to a slightly different constraint. In particular, we first consider the case where the CO and [C{\scriptsize II}] lines originate from different regions of the molecular cloud. This picture is consistent with the standard structure of PDRs in which there is a layer of almost totally ionized carbon at the outer edge, intermediate regions where the carbon is atomic, and internal regions where the carbon is locked into CO \citep{dal, tie, wolf}. In this picture then, the hydrogen is fully molecular in the CO emitting regions. We then consider models in which the CO molecules and C$^+$ ions are uniformly mixed, such that the line emissions originate in gas at the same temperature and density. These models resemble conditions found in UV-opaque cosmic-ray dominated dark cores of interstellar molecular clouds where the chemistry is driven entirely by cosmic-ray ionization. For HDF850.1, the ionization rates could be significantly higher than in the Milky Way, leading to enhanced C$^+$ in the UV-shielded regions. In both instances, we perform our LVG computations assuming \emph{(a)} virialized clouds for which the virialization condition (Equation [11]) has been imposed and \emph{(b)} gravitationally-unbound clouds. 
\\ \indent To carry out these computations, we use the Mark \& Sternberg LVG radiative transfer code described in \citet{dave}. Energy levels, line frequencies and Einstein A coefficients are taken from the Cologne Database for Molecular Spectroscopy (CDMS).The excitation and deexcitation rates of the CO rotational levels that are induced by collisions with H$_2$ are taken from \citet{yang} while the C$^+$ collisional rate coefficients come from \citet{flowla} and \citet{laro}.

\subsection{Separate CO, C$^+$ Virialized Regions}
We first consider a model in which the CO and [C{\scriptsize II}] emission lines detected at the position of HDF 850.1 originate in separate regions of the molecular gas cloud, regions which are not necessarily at the same temperature and number density. For self-gravitating clouds in virial equilibrium, the velocity gradient is no longer an independent input parameter of the LVG model, but varies with $n_{H_2}$ according to equation (11). To find the unique solution that yields the two observed line ratios, $I_{CO(6-5)}/I_{CO(2-1)}$ and $I_{CO(6-5)}/I_{CO(5-4)}$, we assume a canonical value of $\bigchi_{CO}$ = 10$^{-4}$ for the relative CO to H$_2$ abundance and vary the remaining two parameters, temperature and molecular hydrogen density, over a large volume of the parameter space. We find, under this virialization constraint, that the observed CO lines are best fit with a kinetic temperature of 70 K and a molecular hydrogen number density of 10$^{2.6}$ cm$^{-3}$ (with a corresponding velocity gradient of $\approx$ 0.16 km s$^{-1}$pc$^{-1}$). The column density that yields the correct line intensity magnitudes is 4.2$\times$10$^{19}$ cm$^{-2}$, corresponding to a molecular hydrogen gas mass of M$_{H_2}\approx$ 2.13$\times$10$^{11}$ M$_{\odot}$. The H$_2$ mass to CO luminosity conversion factor obtained in this model is $\alpha$ = 5.1 M$_{\odot}$(K km s$^{-1}$ pc$^2$)$^{-1}$, a value similar to the Galactic conversion factor observed for virialized molecular clouds in the Milky Way, suggesting that HDF 850.1 may have some properties in common with our Galaxy. 
\\ \indent Reducing the relative CO to H$_2$ abundance by a factor of two, to $\bigchi_{CO}$ = 5$\times$10$^{-5}$, results in a best fit solution with a molecular hydrogen gas mass of M$_{H_2}\approx$ 2.68$\times$10$^{11}$ M$_{\odot}$, nearly 25\% larger than the value obtained assuming $\bigchi_{CO}$ = 10$^{-4}$. Ranges on the fit parameter consistent with the observational uncertainties are listed in Table 2.

\subsection{Separate CO, C$^+$ Unvirialized Regions}
We then consider a model in which the CO and [C{\scriptsize II}] emission lines are assumed to originate from separate regions of gravitationally-unbound molecular clouds. Since unvirialized clouds generally demonstrate a higher degree of turbulence relative to their virialized counterparts,  we expect the velocity gradient in this model to be greater than the velocity gradient obtained for the virialized model, $(dv/dr)_{virialized} \approx$ 0.16  km s$^{-1}$pc$^{-1}$. We therefore fix the velocity gradient to be ten times the virialized value, $(dv/dr)_{unvirialized}$ = 1.6 km s$^{-1}$pc$^{-1}$, and, assuming a canonical value of $\bigchi_{CO}$ = 10$^{-4}$, find the solution that yields the two observed line ratios by varying $T$ and $n_{H_2}$. We find, under these assumptions, that the CO SED is best fit with a molecular hydrogen density of 10$^3$ cm$^{-3}$ and
a kinetic temperature of 100 K. For this set of parameters, the beam-averaged H$_2$ column density is $N_{H_2} \approx$ 1.0$\times$10$^{19}$ cm$^{-2}$, giving an associated molecular gas mass of $M_{H_2} \approx$ 5.16$\times$10$^{10}$  M$_{\odot}$. This estimate of the gas mass is nearly 50\% larger than that obtained by \citet{walt} by applying the H$_2$ mass-to-CO luminosity 
relation with the typically adopted conversion factor for ULIRGs, $\alpha$ = 0.8 M$_{\odot}$ (K km s$^{-1}$ pc$^2$)$^{-1}$. Given our inferred H$_2$ gas mass and predicted CO(1-0) line luminosity from our LVG fit, we find that $\alpha$ = 1.2 M$_{\odot}$ (K km s$^{-1}$ pc$^2$)$^{-1}$, in this model.
\\ \indent The increase in molecular hydrogen density and the reduction in inferred mass (relative to the values obtained in the previous model where the virialization condition was imposed) arise from our assumption that the velocity gradient in this model is greater than the velocity gradient obtained in the virialized case. For a fixed $\bigchi$, the optical depth drops with increasing $dv/dr$ (Equation [7]); since $\beta$, the probability of an emitted photon escaping, correspondingly increases,  the radiation is less ``trapped" and a higher density, $n_{H_2}$, is required to produce the observed CO excitation lines. Furthermore, since $I_{i,j} \propto\beta_{i,j}M$, a larger $\beta$ implies that less mass is required to reproduce an observed set of line intensities. Therefore, assuming $(dv/dr)_{unvirialized}$ = 10 $(dv/dr)_{virialized}$ causes the optical depth, and consequently, the inferred mass, to drop by a factor of nearly 4 in this model.

\begin{figure*}
\centerline{\includegraphics[width=400pt]{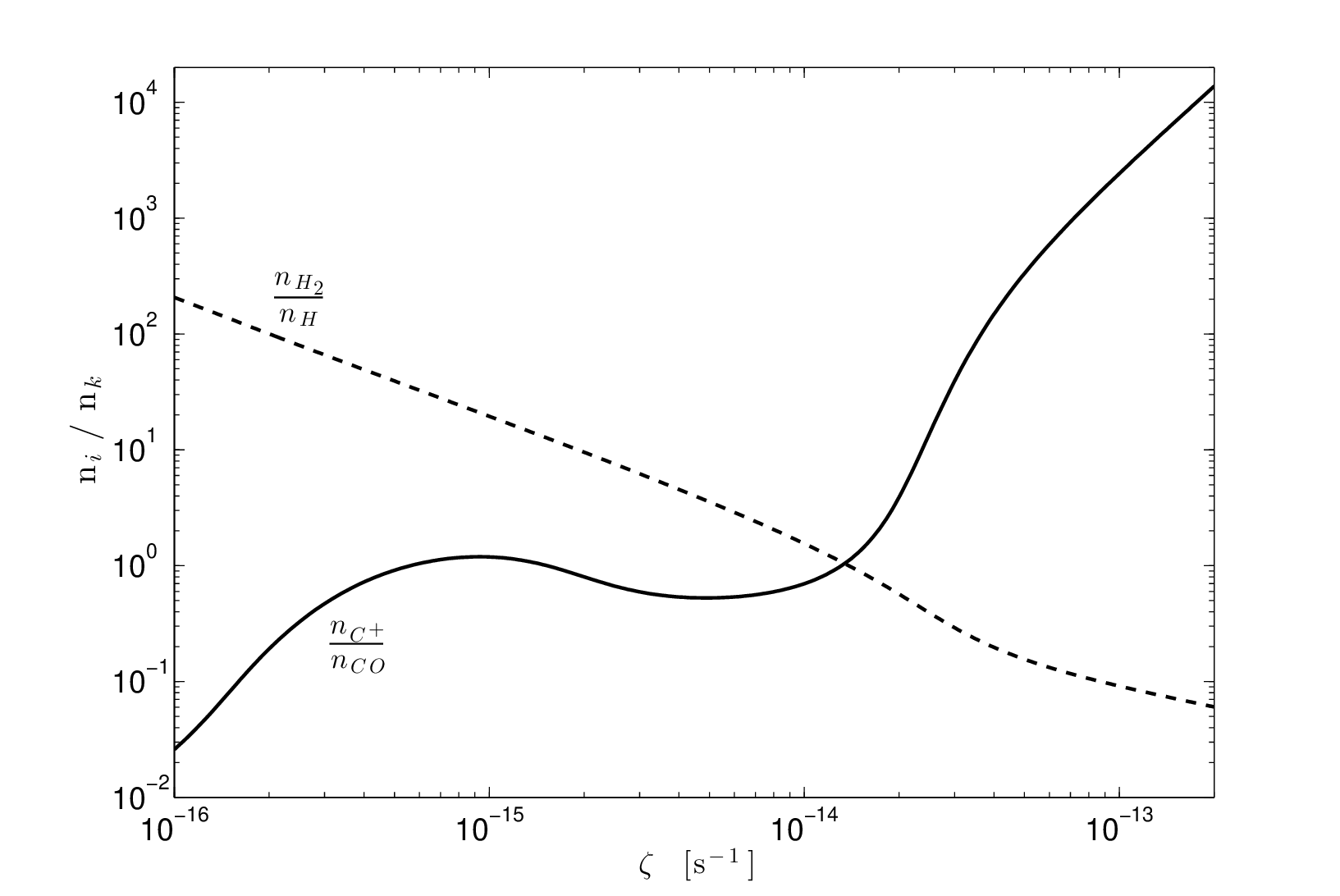}}
\caption{Dependence of the density ratios, $n_{C^+}/n_{CO}$ (solid line) and $n_{H_2}/n_{H}$ (dashed line), on the H$_2$ ionization rate $\zeta$ at a fixed solar metallicity \emph{Z'}=1, for our best-fit LVG parameters T$_{kin}$ = 160 K and $n_{H_2}$ = 10$^{3}$ cm$^{-3}$). As $\zeta$ increases, the abundance of C$^+$ relative to CO in the cosmic-ray dominated dark cores of interstellar clouds grows while that of H$_2$ to atomic hydrogen decreases. The desired value $n_{C^+}/n_{CO}\approx$ 13 is obtained for $\zeta \approx$ 2.5$\times$10$^{-14}$ s$^{-1}$, at which point $n_{H_2} \approx  0.4n_{H}$.}
\end{figure*}

\subsection{Optically thin [C{\scriptsize II}]}
In the two models above, where the CO and [C{\scriptsize II}] lines are assumed to be emitted from separate regions of the molecular clouds, the single detected ionized carbon line is insufficient in constraining the parameters of the LVG modeled [C{\scriptsize II}] region.  We thus consider the optically thin regime of the [C{\scriptsize II}] line ($\beta \simeq$ 1), such that
\begin{equation}
I_{ij}\simeq \frac{h\nu_{ij}}{4\pi}A_{ij}x_i \bigchi_{C^+}N_{H_2}  \rightarrow \frac{M_{H_2}}{L_{[C{\scriptsize II}]_{ij}}}=\frac{\mu m_{H_2}(1+z)^4}{h\nu_{ij}A_{ij}x_i\bigchi_{C^+}}
\end{equation}
where $\bigchi_{C^+}$ is the C$^+$ to H$_2$ abundance ratio (fixed at a value of 10$^{-4}$ in these calculations) and $x_i$ represents the fraction of ionized carbon molecules in the \emph{i}th energy level. Assuming that the fine-structure transition J=3/2$\rightarrow$1/2 is due solely to spontaneous emission processes and collisions with ortho- and para-H$_2$, the equations of statistical equilibrium reduce to a simplified form and can be solved to obtain \emph{x$_{(J=3/2)}$} as a function of temperature and H$_2$ number density. In Fig. 1, we have plotted the resulting mass-to-luminosity ratio, $M_{H_2}/L_{[C{\scriptsize II}]}$, as a function of the kinetic temperature for several different values of $n_{H_2}$. Given the high [C{\scriptsize II}]/far-infrared luminosity ratio of $L_{[C{\scriptsize II}]}/L_{FIR}$ = (1.7$\pm$0.5)$\times$10$^{-3}$ in HDF 850.1 \citep{walt}, it is reasonable to assume the ionized carbon is emitting efficiently and to thus consider the high temperature, large number density limit ($T_{kin}\simeq$ 500 K, $n_{H_2}\simeq$ 10$^4$ cm$^{-3}$). In this limit, the mass-to-luminosity ratio is $\approx$ 1.035 M$_{\odot}$(K km$^{-1}$ pc$^{2})^{-1}$ and the corresponding molecular gas mass of the C$^+$ region, using the detected line intensity of the ionized carbon line $L_{[C{\scriptsize II}]}$ = 1.1$\times$10$^{10}$ L$_{\odot}$, is found to be M$_{H_2}\approx$ 5.2$\times$10$^{10}$ M$_{\odot}$ .

\subsection{Uniformly Mixed CO, C$^+$ Virialized Region}
We next consider models in which the CO molecules and C$^+$ ions are mixed uniformly, such that the corresponding line emissions originate in gas at the same temperature, density, and velocity gradient. For these conditions, the chemistry is driven by cosmic-ray ionization and the density fractions \emph{n$_{i}$/n} for CO and C$^+$ depend on a single parameter, the ratio of the cloud density to the cosmic-ray ionization rate $\zeta$ \citep{boger}. For Galactic conditions with $\zeta\sim$10$^{-16}$ s$^{-1}$, the C$^+$ to CO ratio is generally very small. However, in objects such as HDF 850.1 where the ionization rate may be much larger due to high star formation rates, this ratio may be enhanced significantly.
\\ \indent Assuming a virialized cloud with $\bigchi_{CO}$ + $\bigchi_{C^+}$ = 10$^{-4}$, the unique LVG fit for the observed set of line intensity ratios, \{$I_{[C{\scriptsize II}]}/I_{CO(2-1)}$, $I_{[C{\scriptsize II}]}/I_{CO(5-4)}$, $I_{[C{\scriptsize II}]}/I_{CO(6-5)}$\} = \{7.08$\pm$1.67, 0.96$\pm$0.19, 1.03$\pm$0.2\}$\times$10$^2$ \citep{walt}, yields $T_{kin}$ = 160 K, $n$ = 10$^{3}$ cm$^{-3}$ and a C$^+$ to CO abundance ratio of 13. To estimate the ionization rate required to produce this C$^+$/CO abundance ratio, we employed the \citet{boger} chemical code that captures the C$^+$ -- C -- CO interconversion in a purely ionization-driven chemical medium. 
\\ \indent For solar abundances of the heavy elements ($Z'$ = 1), a fairly reasonable assumption given the high star formation rate observed in HDF 850.1, the cosmic-ray ionization rate required to achieve this high C$^+$ to CO ratio in a cloud with temperature \emph{T}=160 K and density $n_{H_2}$=10$^{3}$ cm$^{-3}$ is of order  $\zeta \simeq$ 2.5$\times$10$^{-14}$ s$^{-1}$ (Figure 2). This is significantly enhanced compared to the Milky Way value, $\zeta \sim$10$^{-16}$ s$^{-1}$, and is plausibly consistent with the fact that HDF 850.1 has a star-formation rate of 850 solar masses per year, a value which is larger than the measured Galactic SFR  by a factor of 10$^{3}$ \citep{walt}. 
\\ \indent Furthermore, for such a high cosmic-ray ionization rate of this magnitude, \emph{the hydrogen is primarily atomic for the implied LVG gas density}. We find that the ratio of molecular hydrogen to atomic hydrogen in the interstellar clouds is n$_{H_2}$/n$_{H}\approx$ 0.4 (Figure 2). We thus replace H$_2$ with H as the dominant collision partner. Given the uncertain CO-H rotational excitation rates (see \citealt{shep}), we assume that the rate coefficients are equal to the rates for collisions with ortho-H$_2$ \citep{flopin}. We find that the observed CO and [C{\scriptsize II}] lines together are best fit with a temperature of 160 K, an atomic hydrogen number density of n$_H$ = 10$^{3}$ cm$^{-3}$ (with a corresponding velocity gradient of 0.17 km s$^{-1}$ pc$^{-1}$), and an abundance ratio (relative to H) of 9.3$\times$10$^{-5}$ and 7$\times$10$^{-6}$ for C$^+$ and CO respectively. The column density that yields the correct line intensity magnitudes is 8.5$\times$10$^{19}$ cm$^{-2}$, corresponding to an atomic hydrogen gas mass of M$_H\approx$ 2.14$\times$10$^{11}$ M$_{\odot}$. The molecular hydrogen mass is then M$_{H_2}$ = 2(n$_{H_2}$/n$_H$)M$_H \approx $ 1.72$\times$10$^{11}$ M$_{\odot}$ and the total gas mass estimate is M$_{gas} \approx$ 3.86$\times$10$^{11}$ M$_{\odot}$. The corresponding conversion factor in this model is $\alpha$ = 9.8 M$_{\odot}$(K km s$^{-1}$ pc$^2$)$^{-1}$, where $\alpha$ is now defined as the \emph{total} gas mass to CO luminosity ratio.
\\ \indent Reducing the fixed sum of abundance ratios (relative to H) of CO and C$^{+}$ by a factor of two, to $\bigchi_{CO}$ + $\bigchi_{C^+}$ = 5$\times$10$^{-5}$, results in a best fit solution with a total gas mass of $M_{gas} \approx$ 4.57$\times$10$^{11}$ M$_{\odot}$, nearly 20\% larger than the value obtained assuming $\bigchi_{CO}$ + $\bigchi_{C^+}$ = 10$^{-4}$.

\subsection{Uniformly Mixed CO, C$^+$ Unvirialized Region}
In the case where we assume gravitationally-unbound molecular clouds, we again find a best fit model with a C$^+$ to CO ratio of $\bigchi_{C^+}/\bigchi_{CO}$ $\approx$ 13, indicating that atomic hydrogen is the dominant collision partner in the LVG calculations.  We thus calculate a three-dimensional grid of model CO and [C{\scriptsize II}] lines, varying $T_{kin}$, $n_{H}$, and the relative abundances $\bigchi_{[C{\scriptsize II}]}$ and $\bigchi_{CO}$, with the constraints that $\bigchi_{CO}$ + $\bigchi_{C^+}$ = 10$^{-4}$ and $(dv/dr)_{unvirialized}$ = 10$(dv/dr)_{virialized}$ = 1.7 km s$^{-1}$ pc$^{-1}$ . The observed set of CO and [C{\scriptsize II}] lines, assumed to have been emitted from the same region, are fit best with a temperature of 180 K, an atomic hydrogen number density of 10$^{3.4}$ cm$^{-3}$ and an abundance ratio of 9.3$\times$10$^{-5}$ and 7$\times$10$^{-6}$ for C$^+$ and CO respectively. For this set of parameters, the beam-averaged H column density is well constrained to be N$_H \approx$ 2.9$\times$10$^{19}$ cm$^{-2}$. This corresponds to an atomic and molecular gas mass of M$_H \approx$ 7.33$\times$10$^{10}$ M$_{\odot}$ and M$_{H_2} \approx$ 5.20$\times$ 10$^{10}$ M$_{\odot}$ respectively, yielding a total gas mass estimate of M$_{gas} \approx$ 1.25$\times$10$^{11}$ M$_{\odot}$. The ratio of the total gas mass to the CO luminosity in this model is $\alpha$ = 5.1 M$_{\odot}$(K km s$^{-1}$ pc$^2$)$^{-1}$.

\begin{table*}
\begin{minipage}{140mm}
  \caption{LVG Model: Best Fit Parameters and Results}
  \centering
  \begin{tabular}{@{}lrrrrrlrlr@{}}
  \hline
   Model Parameters%
    \footnote{For each model parameter, the top row represents the unique, best fit value obtained for the specified model. The bottom row provides the range of parameter values that yield results consistent with the observed line intensity ratios within the error bars of the observed data points \citep{walt}.}
    & Separate, & Separate, & Mixed, & Mixed,\\
   & virialized & unvirialized & virialized & unvirialized\\
   \hline
   $T_{kin}$ [K] & 70 & 100 & 160 & 180\\
   & [30,180] & [30,200] & [120,180] & [140,260]\\
   \\
   $dv/dr$ [km s$^{-1}$ pc$^{-1}$] & 0.16 & 1.6 & 0.17 & 1.7\\
   & [0.14,0.20] & -- & [0.14,0.28] & --\\
   \\
   $\log_{10} n_{H_2}$ [cm$^{-3}$] & 2.6 & 3 & 2.6 & 2.9\\
   & [2.5,2.8] & [3,3.2] & [2.6,2.9] & [2.8,3.1]\\
   \\
   $\log_{10} n_{H}$ [cm$^{-3}$] & -- & -- & 3 & 3.4\\
   & -- & -- & [2.8,3.4] & [3.3,3.5]\\
   \\
   $\bigchi_{CO/H_2}$ & 10$^{-4}$ & 10$^{-4}$ & 7$\times$10$^{-6}$ & 7$\times$10$^{-6}$ \\
   & -- & -- & [5,9]$\times$10$^{-6}$ & [5,9]$\times$10$^{-6}$\\
   \\
   $\bigchi_{CO/H}$ & -- & -- & 9.3$\times$10$^{-5}$ & 9.3$\times$10$^{-5}$\\
   & -- & -- & [9.1,9.5]$\times$10$^{-5}$ & [9.1,9.5]$\times$10$^{-5}$\\
   \\
   $N_{H_2}$ [10$^{19}$ cm$^{-2}$] & 4.2 & 1.0 & 3.4 & 1.0\\
   & [1.9,15] & [0.5,5.4] & [2.0,7.4] & [0.6,1.5]\\
   \\
   $N_H$ [10$^{19}$ cm$^{-2}$] & -- & -- & 8.5 & 2.9\\
   & -- & -- & [6.1,12] & [2.3,3.8]\\
  \hline
   Model Results & & & &\\
   \hline
   $M_{H_2}$ [M$_{\odot}$] & 2.13$\times$10$^{11}$ & 5.16$\times$10$^{10}$ & 1.72$\times$10$^{11}$ & 5.20$\times$10$^{10}$\\
   & [9.26,74.9]$\times$10$^{10}$ & [2.61,27.2]$\times$10$^{11}$ & [1.11,3.73]$\times$10$^{11}$ & [3.37,7.65]$\times$10$^{10}$\\
  \\
   $M_{H}$ [M$_{\odot}$] & -- & -- & 2.14$\times$10$^{11}$ & 7.33$\times$10$^{10}$\\
   & -- & -- & [1.54,3.01]$\times$10$^{11}$ & [5.85,9.4]$\times$10$^{10}$\\ 
   \\
   $M_{gas}$ [M$_{\odot}$] & 2.13$\times$10$^{11}$ & 5.16$\times$10$^{10}$ & 3.86$\times$10$^{11}$ & 1.25$\times$10$^{11}$\\
   & [9.26,74.9]$\times$10$^{10}$ & [2.61,27.2]$\times$10$^{11}$ & [2.53,6.74]$\times$10$^{11}$  & [9.22,17.0]$\times$10$^{10}$\\
   \\ 
   $L_{CO(1-0)}$ [(K km s$^{-1}$ pc$^{2}$)] & 3.95$\times$10$^{10}$ & 4.41$\times$10$^{10}$ & 2.49$\times$10$^{10}$ & 4.18$\times$10$^{10}$\\
   \\
    $\alpha$%
    \footnote{$\alpha$ here is defined as the ratio of total gas mass to CO(1-0) luminosity, $\alpha$ = M$_{gas}$/L'$_{CO(1-0)}$. In the models where the CO and [C{\scriptsize II}] lines are assumed to be originating from separate regions, M$_{gas}$ = M$_{H_2}$ since estimates of the atomic hydrogen gas mass could not be obtained via the LVG calculations. In the models where the CO molecules and C$^+$ ions  are assumed to be uniformly mixed, the total gas mass is the sum of the molecular and the atomic gas masses, M$_{gas}$ = M$_{H_2}$+M$_H$.}
 [M$_{\odot}$(K km s$^{-1}$ pc$^{2}$)$^{-1}$] & 5.1 & 1.2 & 9.8 & 5.1\\
    \end{tabular}
\end{minipage}
\end{table*}

 \section{Cosmological Constraints}
Our inferred gas masses enable us to set cosmological constraints. For a particular set of cosmological parameters, the number density of dark matter halos of a given mass can be inferred from the halo mass function. The Sheth-Tormen mass function, which is based on an ellipsoidal collapse model, expresses the comoving number density of halos \emph{n} per logarithm of halo mass \emph{M} as,
 \begin{equation}
n_{ST}(M)=\frac{dn}{d logM}=A\sqrt{\frac{2a}{\pi}}\frac{\rho_m}{M}\frac{d\nu}{d logM}(1+\frac{1}{(a\nu^2)^p})e^{-a\nu^2/2}
\end{equation}
where a reasonably good fit to simulations can be obtained by setting \emph{A} = 0.322, \emph{a} = 0.707, and \emph{p} = 0.3 \citep{sheth}. Here, $\rho_{m}$ is the mean mass density of the universe and $\nu$ = $\delta_{crit}(z)/\sigma(M)$ is the number of standard deviations away from zero that the critical collapse overdensity represents on mass scale \emph{M}. Integrating this comoving number density over a halo mass range and volume element thus yields \emph{N}, the expectation value of the total number of halos observed within solid angle $A$ with mass greater than some \emph{M$_h$} and redshift larger than some $z$,
\begin{align}
&N(z, M_{h})= \nonumber\\
&\frac{c}{H_0}A \int_{z}^{\infty}\!\!\!dz\frac{D_A(z)^2 }{\sqrt{\Omega_m(1+z)^3+\Omega_{\Lambda}}}\int_{M_{h}}^{\infty}\!\!\!\!dlogM \frac{dn}{dlogM} .
\end{align}
where $H_0$ is the Hubble constant, $D_A$ is the angular diameter distance, and $\Omega_m$ and $\Omega_\Lambda$ are the present-day density parameters of matter and vacuum, respectively.

Under the assumption that the number of galaxies in the field of observation follows a Poisson distribution, the probability of observing at least one such object in the field is then $P$ = $1-F(0,N)$ where $F(0,N)$ is the Poisson cumulative distribution function with a mean of $N$. Given the detection of HDF 850.1, we can say that out of the hundreds of submillimetre-bright galaxies identified so far, at least one has been detected in the Hubble Deep Field at a redshift z $>$ 5 with a halo mass greater than or equal to the halo mass associated with this source. This observation, taken together with the theoretical number density predicted by the Sheth-Tormen mass function, implies that an atomic model that yields an expectation value $N$ can be ruled out at a confidence level of 

\begin{equation}
F(0,N(5, M_{h,min}))
\end{equation}
where the solid angle covered by the original SCUBA field in which HDF850.1 was discovered is $\simeq$ 9 arcmin$^2$ \citep{hughes} and a $\Lambda$CDM cosmology is assumed with H$_0$ = 70 km s$^{-1}$ Mpc$^{-1}$, $\Omega_\Lambda$ = 0.73, and $\Omega_m$ = 0.27 \citep{kom}. $M_{h,min}$, the minimum inferred halo mass for HDF 850.1, is related to the halo's minimum baryonic mass component, a quantity derived in \S3 via the LVG technique, in the following way,  
\begin{equation}
M_{h,min} = \frac{\Omega_m}{\Omega_b} M_{b,min}
\end{equation}
where the baryonic and the total matter density parameters are  $\Omega_b$ = 0.05 and $\Omega_m$ = 0.27 respectively. Each model's estimate of the minimum baryonic mass associated with HDF 850.1 therefore corresponds to an estimate of \emph{N(5,M$_{h,min}$)} and respectively yields the certainty with which the model can be discarded.
\\ \indent The confidence with which models can be ruled out on this basis is plotted as a function of the minimum baryonic mass estimated by the model (solid curve in Figure 3). To check the consistency of the four LVG models considered in \S3 with these results, dashed lines representing the masses derived from each model are included in the plot (upper left panel). Assuming the CO and [C{\scriptsize II}] molecules are uniformly mixed in virialized clouds results in a baryonic mass M$_{b,min}$ =  3.86$\times$10$^{11}$ M$_{\odot}$. The probability of observing at least one such source, with a corresponding halo mass M$_{h} \geq$ 2.1$\times$10$^{12}$ M$_{\odot}$, is $\sim$ 7$\times$10$^{-2}$; this model can thus be ruled out at the $1.8\sigma$ level (solid). 
The model which postulate separate virialized regions (dashed) can be ruled out with relatively less certainty, at the $1\sigma$ level. On the other hand, modeling the CO and [C{\scriptsize II}] emission lines as originating from mixed (dotted) or separate (dash-dot) unvirialized regions, results in minimum baryonic masses which are consistent with the constraint posed by equation (15). We expect to find $N \sim$ 1.5 and 10 such halos, respectively, at a redshift z $\geq$ 5. The fact that the 
 expected average number of observed sources for the latter model is much higher than the actual number of sources observed may be due to the incompleteness of the conducted survey and is therefore not grounds for ruling out this model. 
\\ \indent The baryonic masses, obtained via the LVG method, represent conservative estimates of the total baryonic content associated with HDF 850.1. In particular, mass contributions from any ionized gas or stars residing in the galaxy are not taken into account, and in the models where a layered cloud structure is assumed, the atomic hydrogen mass component is left undetermined. Furthermore, ejection of baryons to the IGM through winds may result in a halo baryon mass fraction that is smaller than the cosmic ratio, $\Omega_b/\Omega_m$, used in this paper, resulting in a conservative estimate of the minimum halo mass. We therefore consider the effects on the predicted number of observed halos, \emph{N(M$_{h,min}$)}, if the minimum baryonic mass derived for each model is doubled, e.g. assuming a molecular-gas mass fraction of $\sim$ 1/2 \citep{tacc2013}. Using these more accurate estimates of HDF 850.1's baryonic mass, we find that the two models in which the virialization condition is enforced can be ruled out at the $\sim 2.7\sigma$ and $2\sigma$ level for the mixed and separate models, respectively (upper right panel).
\\ \indent For comparison, we also consider the gas masses implied by the CO(1-0) line luminosities predicted by the two models where distinct CO and C$^+$ layers are assumed (lower left panel). Adopting a CO-to-H$_2$ conversion factor of $\alpha$ = 0.8 (K km s$^{-1}$ pc$^2$)$^{-1}$, we find that enforcing  the cosmological constraint posed by the abundance of dark matter halos does not rule out either of the two models in which separate CO and C$^+$ regions were assumed, even if the minimum baryonic masses are doubled to account for neglected contributions to the total gas mass (lower right panel). If a conversion factor of $\alpha$ = 4.6 (K km s$^{-1}$ pc$^2$)$^{-1}$ is used, both models can be ruled out at the $\sim$ 1$\sigma$ level and increasing these obtained masses by a factor of two drives up the sigma levels to $\sim1.8\sigma$ for both models.

 \begin{figure*}
\subfloat {\hspace{-1.3 cm} \includegraphics[width=310pt]{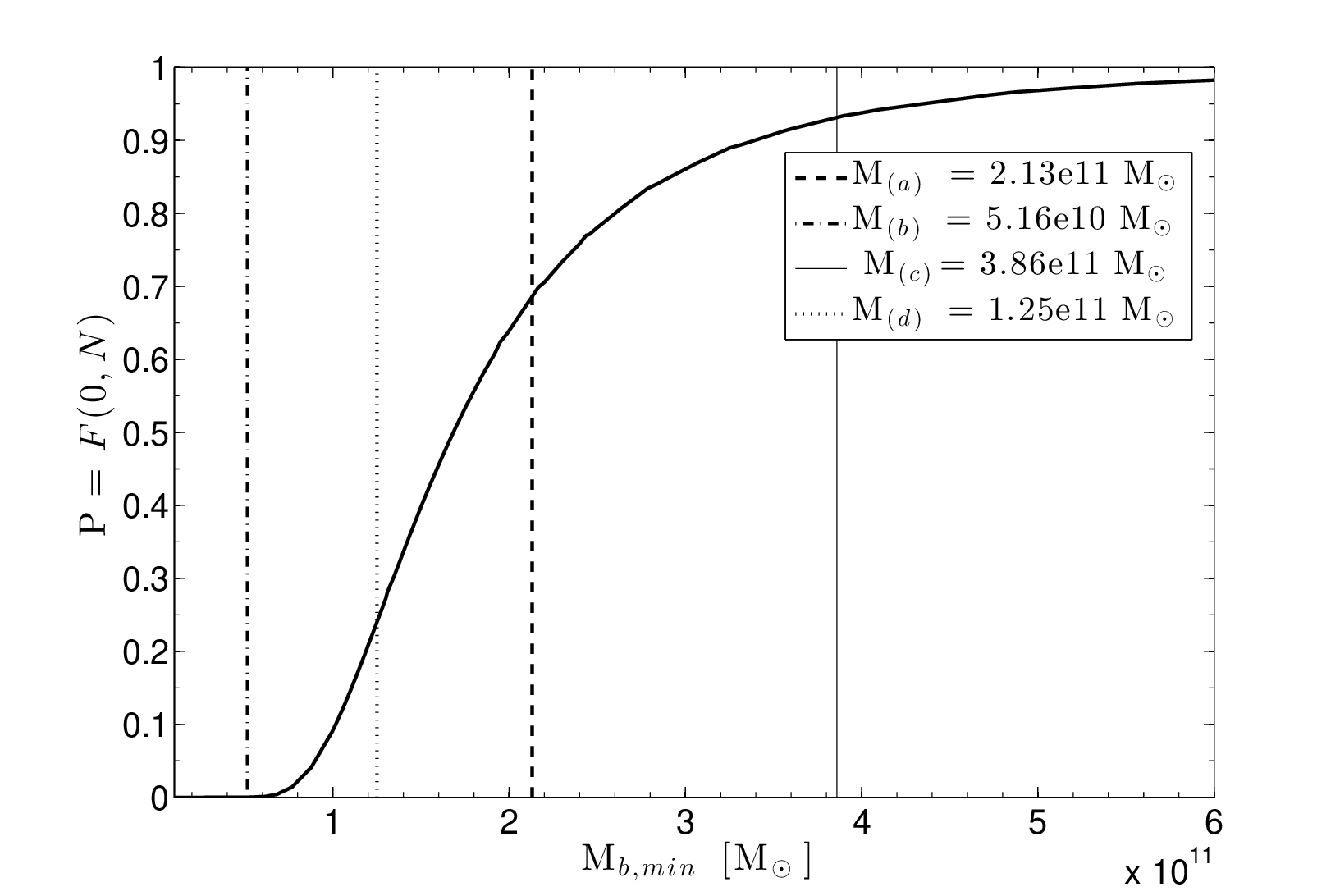}}
\subfloat{\hspace{-1 cm}\includegraphics[width=310pt]{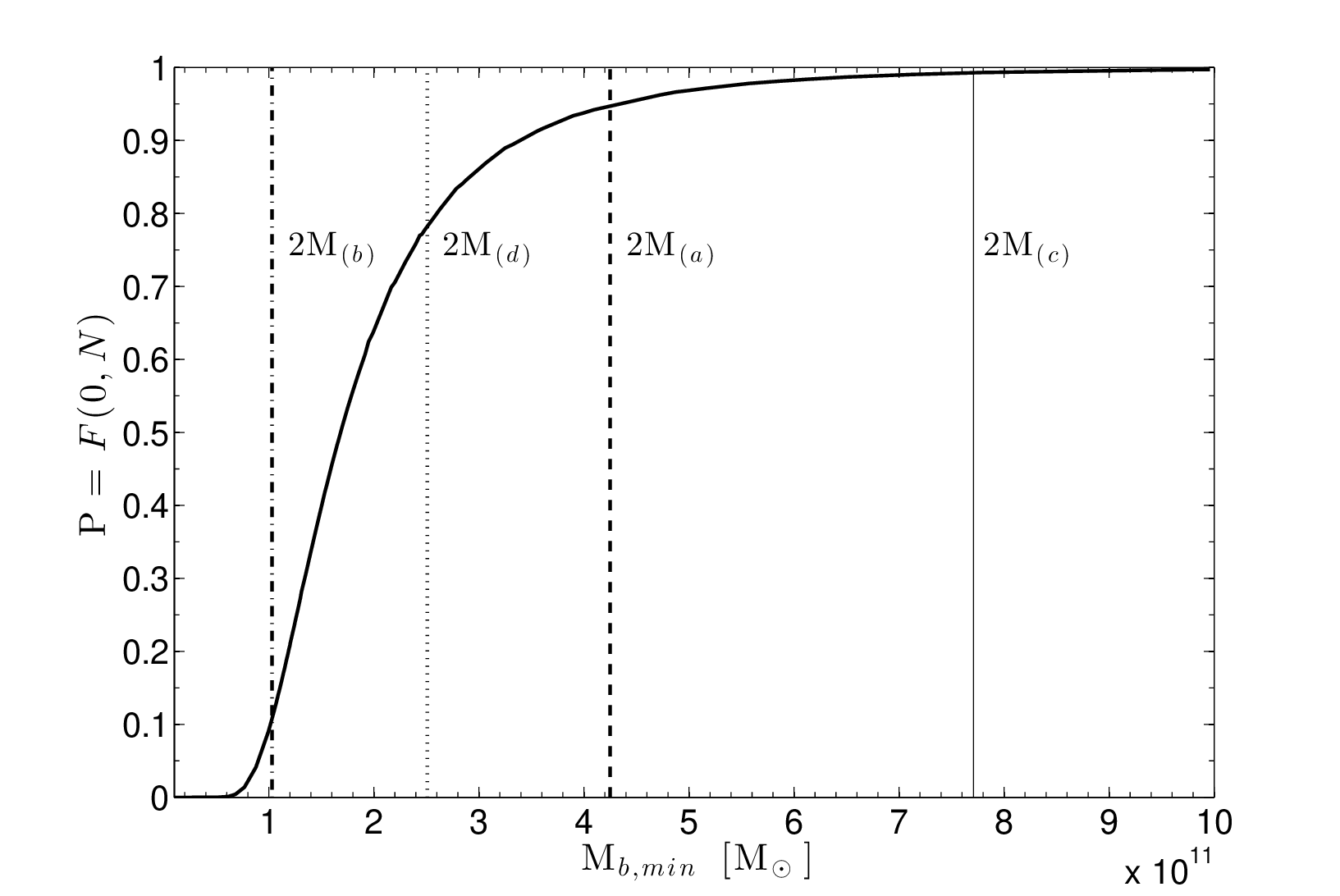}}\\
\subfloat{\hspace{-1.3 cm}\includegraphics[width=310pt]{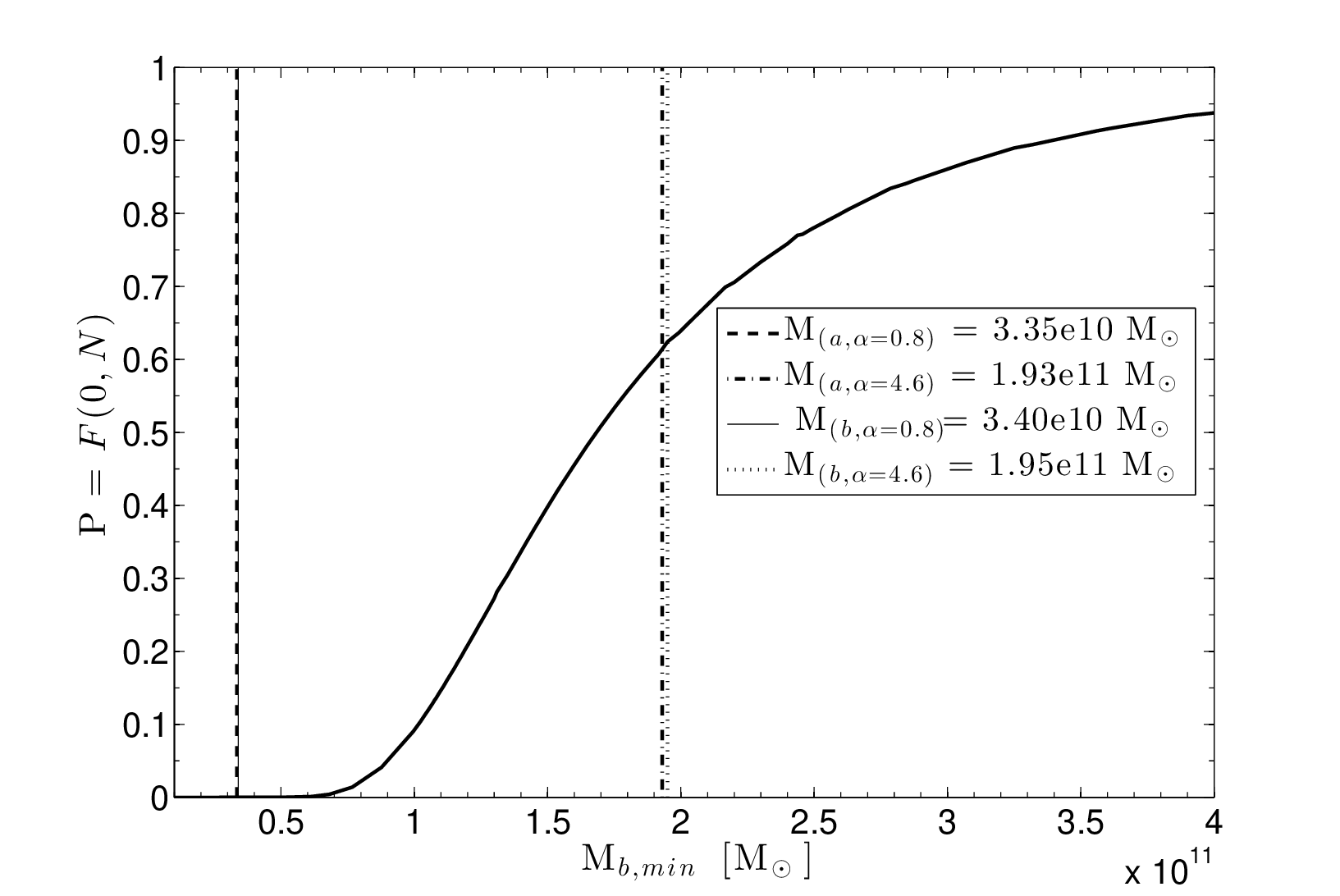}}
\subfloat{\hspace{-1 cm}\includegraphics[width=310pt]{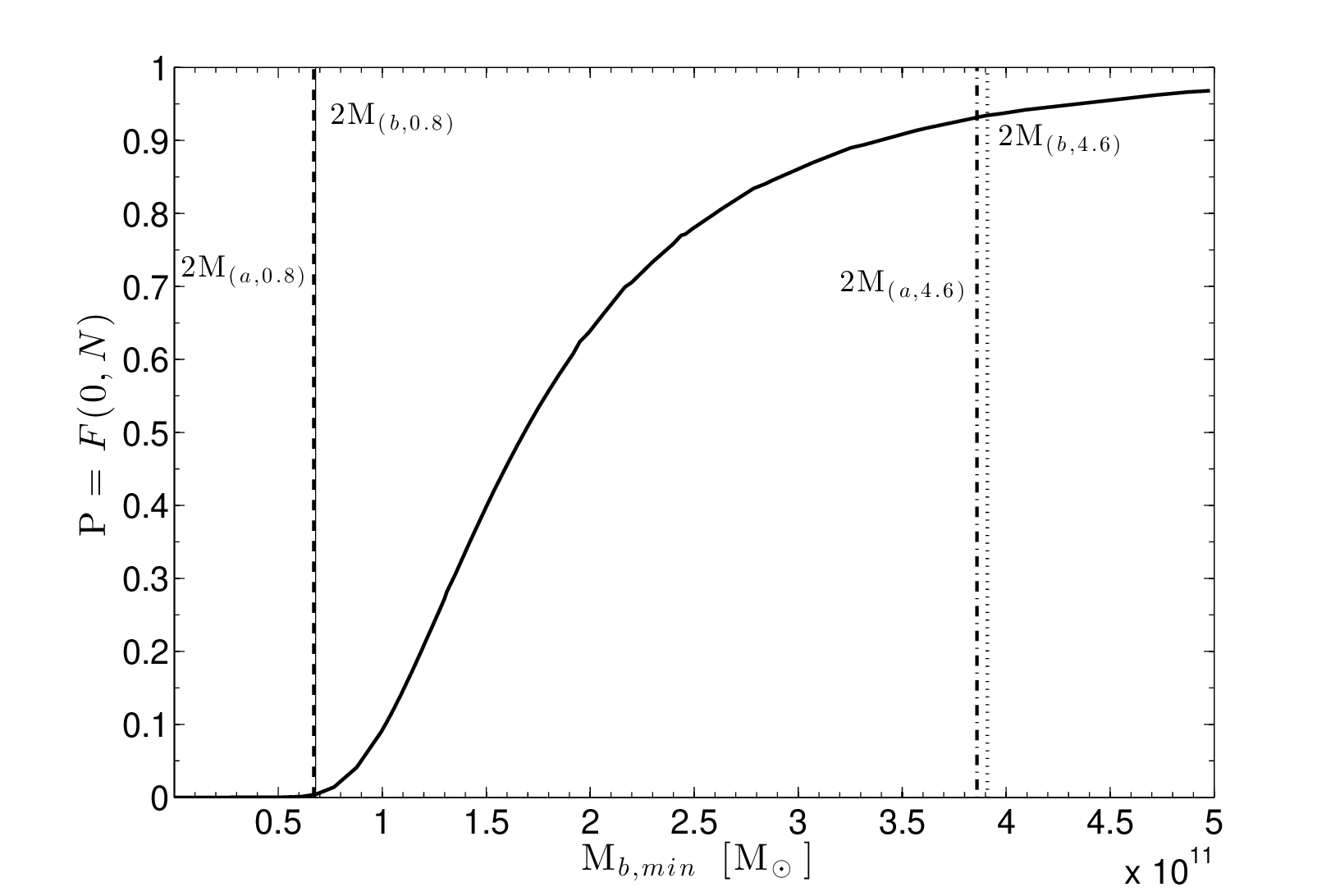}}
 \caption{\emph{N} represents the expectation value of the number of halos one expects to find with mass M$_h \geq $ M$_{b,min}$/f$_b$ at a redshift z $\geq$ 5 within a solid angle of $A_{HDF} \simeq$ 9 arcmin$^2$. Assuming that the number of galaxies in a field of observation follows a Poisson distribution, the probability of observing at least one such object in the field with M$_h \geq $ M$_{b,min}$/f$_b$ at a redshift z $\geq$ 5 is $1$ - $F(0,N(z,M_{h,min}))$ where $F(0,N)$ is the Poisson cumulative distribution function with a mean of $N$. The confidence with which an LVG model can be ruled out as a function of the minimum baryonic mass derived from the model is therefore $P$ = $F(0,N)$ (solid curve). \emph{Upper left panel}: The vertical lines represent the mass values obtained for each of the four models presented in \S3: \emph{(a)} separate and virialized (dashed), \emph{(b)} separate and unvirialized (dash-dot), \emph{(c)} mixed and virialized (solid), and \emph{(d)} mixed and unvirialized regions (dotted). In models \emph{(a)} and \emph{(b)}, M$_{b,min}$ = M$_{H_2}$ while in models \emph{(c)} and \emph{(d)},  M$_{b,min}$ = M$_{H_2}$+M$_H$. \emph{Upper right panel}: The minimum baryonic masses obtained for each model were doubled to account for neglected contributions to the total gas mass; models \emph{(a)} and \emph{(c)} can now be ruled out at the $2\sigma$ and $2.7\sigma$ levels respectively. \emph{Lower left panel}: The vertical lines represent the mass values implied by the predicted CO(1-0) line luminosities from models \emph{(a)} and \emph{(b)} with a CO-to-H$_2$ conversion factor of $\alpha$ = 0.8 and 4.6 M$_{\odot}$ (K km s$^{-1}$ pc$^2$)$^{-1}$. If these minimum baryonic mass values are then doubled (\emph{lower right panel}), both models can be ruled out at the $\sim 1.8\sigma$ level in the case where $\alpha$ = 4.6 M$_{\odot}$ (K km s$^{-1}$ pc$^2$)$^{-1}$ is adopted as the conversion factor.}
 \end{figure*}

\section{Summary}
 In this paper, we employed the LVG method to explore alternate model configurations for the CO and C$^+$ emission lines regions in the high-redshift source HDF 850.1. In particular, we considered emissions originating from \emph{(i)} separate virialized regions, \emph{(ii)} separate unvirialized regions, \emph{(iii)} uniformly mixed virialized regions, and \emph{(iv)} uniformly mixed unvirialized regions. For models \emph{(i)} and  \emph{(ii)} where separate CO and C$^+$ regions were assumed, the kinetic temperature, $T_{kin}$, and the molecular hydrogen density, $n_{H_2}$, were fit to reproduce the two observed line ratios, $I_{CO(6-5)}/I_{CO(2-1)}$ and $I_{CO(6-5)}/I_{CO(5-4)}$, for a fixed canonical value of the CO abundance (relative to H$_2$), $\bigchi_{CO}$ = 10$^{-4}$. The column density of molecular hydrogen, $N_{H_2}$, was then fit to yield the correct line intensity magnitudes and the molecular gas mass was derived for each respective model. In models \emph{(iii)} and  \emph{(iv)} where the CO molecules and C$^+$ ions were assumed to be uniformly mixed with abundance ratios that satisfied the constraint,  $\bigchi_{CO}$ +  $\bigchi_{C^+}$ = 10$^{-4}$, we found that a relatively high ionization rate of  $\zeta \simeq$ 2.5$\times$10$^{-14}$ s$^{-1}$  is necessary to reproduce the set of observed line ratios,  \{$I_{[C{\scriptsize II}]}/I_{CO(2-1)}$, $I_{[C{\scriptsize II}]}/I_{CO(5-4)}$, $I_{[C{\scriptsize II}]}/I_{CO(6-5)}$\}. Since the hydrogen in a cloud experiencing an ionization rate of this magnitude is primarily atomic, two additional parameters, $n_H$ and $N_H$, were introduced and the set of LVG parameters, \{\emph{T$_{kin}$, n$_{H_2}$, $n_H$, $\bigchi_{CO}$, $\bigchi_{C^+}$, N$_{H_2}$}, $N_H$\}, were fit and used to obtain both a molecular and an atomic hydrogen gas mass for each model. The gas masses derived by employing the LVG technique thus represent conservative estimates of the minimum baryonic mass associated with HDF 850.1.
\\ \indent These estimates were then used, together with the Sheth-Tormen mass function for dark matter halos to calculate the average number of halos with mass M$_h \geq$ M$_{b,min}$ that each model predicts to find within the HDF survey volume. Given that at least one such source has been detected, we found that models \emph{(i)} and \emph{(iii)} can be ruled out at the 1$\sigma$ and 1.8$\sigma$ levels respectively. The confidence with which these models are ruled out increases if a less conservative estimate of the baryonic mass is taken; increasing the LVG-modeled gas masses by a factor of two to account for neglected contributions to the total baryonic mass, drives up these sigma levels to $\sim$ 2$\sigma$ and 2.7$\sigma$ respectively. Furthermore, model \emph{(iv)} can now be ruled out at the 1$\sigma$ level as well. We are therefore led to the conclusion that HDF 850.1 is modeled best by a collection of unvirialized molecular clouds with distinct CO and C$^+$ layers, as in PDR models. The LVG calculations for this model yield a kinetic temperature of 100 K, a velocity gradient of 1.6 km s$^{-1}$ pc$^{-1}$, a molecular hydrogen density of 10$^{3}$ cm$^{-3}$, and a column density of 10$^{19}$ cm$^{-2}$. The corresponding molecular gas mass obtained using this LVG approach is M$_{H_2} \approx$ 5.16$\times$10$^{10}$ M$_{\odot}$. For this preferred model we find that the CO-to-H$_2$ luminosity to mass ratio is $\alpha$ = 1.2 (K km s$^{-1}$ pc$^2$)$^{-1}$, close to the value found for ULIRGs in the local universe.

 \section{Acknowledgements}
 We thank Dean Mark for his assistance with the LVG computations. This work was supported by the Raymond and Beverly Sackler Tel Aviv University-Harvard/ITC Astronomy Program. A.L. acknowledges support from the Sackler Professorship by Special Appointment at Tel Aviv University. This work was also supported in part by NSF grant AST-0907890 and NASA grants NNX08AL496 andNNA09DB30A (for A.L.).

\label{lastpage}

\end{document}